\documentstyle[prl, aps, amssymb]{revtex}
\begin{document}
\draft
\wideabs{
\title{$^{13}$C NMR investigation of the superconductor MgCNi$_{3}$ up to 
800K}
\author{P.M. Singer,$^{1}$ T. Imai,$^{1}$ T. He,$^{2}$ M.A. Hayward,$^{2}$ 
and R.J. Cava$^{2}$}
\address{Department of Physics and Center for Materials Science and 
Engineering, M.I.T., Cambridge, MA 02139$^{1}$ }
\address{Department of Chemistry and Princeton Materials
Institute, Princeton University, Princeton, NJ 08544$^{2}$ }
\date{\today}
\maketitle
\begin{abstract}
We report $^{13}$C NMR characterization of the new superconductor MgCNi$_{3}$ 
(He et al., Nature {\bf 411}, 54 (2001)).  We found that both the uniform spin 
susceptibility and the spin fluctuations show a strong enhancement with 
decreasing temperature, and saturate below $\sim$50K and $\sim$20K 
respectively.
The nuclear spin-lattice relaxation rate $1/^{13}T_{1}$ exhibits
typical behaviour for isotropic s-wave superconductivity with a 
coherence peak below $T_{c}=7.0$K that grows with decreasing magnetic 
field.  
\end{abstract}
}

Over the past several years, two new major trends have emerged in 
research on superconductivity. One is the search for superconducting
ground states near magnetic instabilities.  
Lonzarich et al., for example, have demonstrated that application of 
high pressure transforms the antiferromagnetic ground states of some heavy 
Fermion materials into superconductors\cite{Lonzarich}.  The exotic 
superconductor Sr$_{2}$RuO$_{4}$\cite{Maeno,Ishida} exhibits 
more subtle signatures of magnetic correlations in the normal state 
above $T_{c}$\cite{Imai,Sidis}. Another emerging trend is 
the search for superconductivity in intermetallic compounds with light 
elements. Recently Akimitsu et al. discovered superconductivity in 
MgB$_{2}$\cite{MgB2}.  
High frequency phonons induced by the light element, B, are believed to be essential in 
yielding the high $T_{c}$.  
 
A very recent addition to the list of new superconductors along these 
trends is MgCNi$_{3}$ \cite{He,Hayward,Li,Huang,Mao,Ren}. MgCNi$_{3}$ forms a three-dimensional perovskite 
structure.  Mg, C, and Ni replace Sr, Ti, and O in 
SrTiO$_{3}$, respectively.  Six Ni atoms at the face-centered position of each cubic unit cell 
form a three-dimensional network of Ni$_{6}$-octahedra.  Each C atom is located 
in the body-centered position surrounded by a Ni$_{6}$-octahedron cage. 
Ni 3d orbitals and C 2p orbitals form two electronic bands at the Fermi level, one being 
electron-like and the other hole-like\cite{Band1,Band2,Band3,Band4}. 
The superconducting transition temperature is modestly high, 
$T_{c}=7.0$K, for an intermetallic system.  The discovery of a
superconducting ground state in MgCNi$_{3}$ poses 
two interesting questions: Firstly, what is the role played by 
Ni 3d electrons?  Unfilled Ni 3d orbitals usually form magnetically 
correlated bands, 
and recent band calculations indicate predominantly Ni 3d character at 
the Fermi level \cite{Band1,Band2,Band3,Band4}. Furthermore, if C atoms 
are removed from MgCNi$_{3}$, MgNi$_{3}$ is expected to have 
a magnetic ground state \cite{Band2}. These calculations
imply that MgCNi$_{3}$ may also be on the verge of a magnetic 
instability.  Unfortunately, the volatility of Mg in chemical 
reaction processes \cite{He} makes growth of bulk samples difficult, and no 
details of possible electronic correlation effects have been addressed by conventional bulk 
measurement techniques. Secondly, does the presence 
of C, a light element, imply that the phonon-mediated BCS mechanism is at 
work for stabilizing the superconducting ground state?  
Recent tunneling measurements by Mao et al. detected a zero bias 
conductance peak below $T_{c}$\cite{Mao} suggesting otherwise. The tunneling result may be interpreted as an 
indication of the unconventional character of the superconducting pairing 
state.

In this Letter, we report the first $^{13}$C NMR investigation of 
MgCNi$_{3}$ both above and below $T_{c}$. Our sample was 
prepared in a manner similar to that previously reported \cite{He}, but with $^{13}$C 
enriched graphite as the starting material to enhance the intensity of 
the $^{13}$C NMR signal. This allowed us to follow the temperature dependence of 
the $^{13}$C NMR properties in a broad temperature range between 1.7K 
and 800K.  We demonstrate that the normal state properties of 
MgCNi$_{3}$ show the signature of modest electronic correlation effects. 
Both the uniform spin susceptibility $\chi'(q=0)$, as measured by the
Knight shift $^{13}K$, and the spin fluctuations, as 
measured by the nuclear spin-lattice relaxation rate $1/^{13}T_{1}$ divided by 
temperature $T$ ($1/^{13}T_{1}T$), increase monotonically below 800K 
down to $\sim50$K and $\sim20$K respectively, before saturating to 
a constant value. A single {\it Korringa relation} for a Fermi-liquid 
$1/T_{1}TK^{2}=$const., can {\it not} account for the temperature dependences 
of $^{13}K$ and $1/^{13}T_{1}T$ between 800K and $T_{c}$. The most likely 
scenario is that electronic correlation effects enhance both the static 
and dynamic magnetic susceptibility, and a
modestly mass-enhanced Fermi-liquid-like state is realized somewhat above $T_{c}$.  
Over all, the normal state NMR properties of 
MgCNi$_{3}$ bear significant qualitative similarities with those of 
the exotic superconductor Sr$_{2}$RuO$_{4}$.  However, we demonstrate 
that our NMR data below $T_{c}$ are consistent 
with conventional s-wave pairing.    

In Fig.1(a), we present the Fourier transformed $^{13}$C NMR lineshape
from MgCNi$_{3}$ powder. Above $T_{c}$, 
$^{13}$C in MgCNi$_{3}$ has a large, temperature dependent, positive 
NMR Knight shift $^{13}K$ 
with much narrower linewidth than the overall 
NMR shift. The narrow linewidth above $T_{c}$ is consistent with the 
fact that the carbon site has cubic point symmetry which insures
the absence of any shift anisotropy.

In Fig.2(a), we present the temperature dependence of the powder averaged 
NMR Knight shift $^{13}K$.  According to band 
calculations\cite{Band1,Band2,Band3,Band4}, MgCNi$_{3}$ has two bands 
arising from Ni 3d and C 2p hybridization at the Fermi energy.  In terms 
of the spin susceptibility 
$\chi_{j}'(q=0)$ of the j-th band ($j=1, 2$ is the band index), one can write $^{13}K$ as \cite{Yafet}
\begin{equation}
          ^{13}K= \sum_{j} A_{j} \chi_{j}'(q=0) + ^{13}K_{orb}.
\label{T1}
\end{equation}
where $A_{j}$ is the hyperfine coupling constant between the $^{13}$C 
nuclear spin ($I=\frac{1}{2}$) and the electrons in the j-th band. 
$^{13}K_{orb}$ is the orbital contribution in the nearly filled 2p orbitals 
of the C atoms and in various materials such as graphite, it is known to be as 
small as 0.01$-$0.02\%.
$A_{j} \chi_{j}'(q=0)$ represents the spin contribution to the Knight 
shift from the j-th band.  Since the hyperfine interaction is dominated by s-electrons, it is safe to 
assume that $A_{1}\sim A_{2}$. Accordingly, the results in 
Fig.2(a) suggest that the total spin susceptibility 
$\sum_{j}\chi_{j}'(q=0)$ increases by approximately 
55-70\% below 800K down to $T_{c}$, if we take
$^{13}K_{orb}=0.01- 0.02\%$.  Needless to say, we cannot rule out 
the possibility that $\chi_{1}'(q=0)$ and $\chi_{2}'(q=0)$ exhibit
somewhat different temperature dependences. The temperature dependence of $^{13}K$ changes 
curvature at about 120K from positive to negative, and saturates below about $T^{*}\sim$50K. The 
electrical resistivity data also show a change of curvature in the same 
temperature range, and satisfy $\rho \sim T^{n}$  with $n \sim 1.8$ below 
$T^{*}\sim$50K \cite{Ong}. These results 
suggest that an electronic crossover takes place near 
$T^{*}\sim$50K prior to the superconducting transition at 
$T_{c}=7.0 $K.

From standard K$-\chi$ analysis $\cite{Yafet}$, where we choose $^{13}K_{orb}=0.014\%$ from the 
insert to Fig.2(a), we obtained $A_{1} = A_{2} \sim 14$ kOe/$\mu_{B}$, the sum of the diamagnetic and 
Van-Vleck contribution as $\chi_{dia}+\chi_{v.v.} \sim -1.55 \times 
10^{-4}$[e.m.u./mol-f.u.], and the saturated value of $\chi_{spin} \sim 4.77 \times 
10^{-4}$[e.m.u./mol-f.u.] below 50K.
Using $N(E_{f})=4.99$[states/ eV f.u.] $\cite{Band3}$ implies 
that the enhancement of the spin susceptibility $\chi_{spin}$ over the 
band value $\chi_{band}=1.61 \times 10^{-4}$[e.m.u./mol-f.u.] is 
$\chi_{spin}/\chi_{band} \sim 3.0$. This is to be compared with the 
specific heat enhancement $\gamma/\gamma_{band} \sim 2.6$ $\cite{Band3}$. 
These estimations give the Wilson ratio $R_{W} =  \frac 
{\chi_{spin}/\chi_{band}} {\gamma/\gamma_{band}}\sim 1.15$ (see also 
\cite{Hayward}). $R_{W}$=2 is expected in the strongly correlated 
limit, while $R_{W}$=1 in the uncorrelated case, therefore $R_{W}$=1.15 
indicates that the electrons are in a mildly correlated 
state. 

In Fig.2(b), we present the temperature dependence of the $^{13}$C nuclear 
spin-lattice relaxation rate $1/^{13}T_{1}$ divided by temperature 
$T$, $1/^{13}T_{1}T$.  See Fig.1(b) for an example 
of nuclear spin recovery after saturation.
Theoretically, the spin contribution to $1/^{13}T_{1}T$ may be written as the wave vector ${\bf q}$ 
summation of the imaginary part of the dynamical electron spin 
susceptibility $\chi''({\bf q},\omega_{n})$,
 \begin{equation}
          \frac{1}{T_{1}T} = \frac{\gamma_{n}^{2}k_{B}}{\mu_{B}^{2} \hbar} 
	\sum_{i,j}\sum_{{\bf q}}|A_{ij}({\bf q})|^{2} \frac{\chi_{i,j}''({\bf q},\omega_{n})}{\omega_{n}}
\label{T1}
\end{equation}
where $\gamma_{n}=10.7054$ MHz/Tesla is the nuclear gyro-magnetic ratio of 
$^{13}$C, $\omega_{n}$ is the resonance frequency, and $i, j$ are band 
indices \cite{Yafet,Moriya,Walstedt}.  As emphasized 
earlier by Walstedt\cite{Walstedt}, cross terms between different 
bands can exist for the spin-lattice relaxation process, while such cross terms 
do not exist in NMR Knight shifts\cite{Yafet}.  This makes 
separation of various contributions to $1/T_{1}T$ a non-trivial 
matter in multi-band systems such as MgCNi$_{3}$ and Sr$_{2}$RuO$_{4}$.    

The most striking aspect of the $1/^{13}T_{1}T$ data is the continuous 
increase of its magnitude all the way from 800K to about 20K through $T^{*}\sim 50$K.
Our results indicate that spin fluctuations are nearly a factor 3 enhanced with decreasing temperature.  
Within a simple Fermi liquid picture 
ignoring all the complications from the 
multi-band effects mentioned above (i.e. we assume that the two sets of d-p bands exhibit 
identical temperature dependences of spin susceptibility), 
a {\it modified Korringa law}, $1/T_{1,spin}TK_{spin}^{2}=1/S\beta$ should 
hold\cite{Moriya,Narath}.  Here $1/T_{1,spin}$ and $K_{spin}$ are the spin 
contribution, and $S=(\frac{\hbar}{4\pi 
k_{B}})(\frac{\gamma_{e}}{\gamma_{n}})^{2}=4.17\times 10^{-6}$ sec K 
for $^{13}$C.  $\beta$ is a 
quantity that signifies the effects of electronic 
correlations.  
In this scenario, the observed 50\% to 70\% increase in the spin contribution 
to the Knight shift $^{13}K$ between 800K and 120K implies that 
$1/^{13}T_{1}T$ would also increase by a factor $1.5^{2}(=2.3)$ to 
$1.7^{2}(=2.9)$ by the Korringa process from electron-hole pair 
excitations at the Fermi level.  This is consistent with the 
factor 2.6 increase of $1/^{13}T_{1}T$ in the same temperature 
range.  In fact, as shown in the inset to Fig.2(a), above 120K we can fit 
$1/^{13}T_{1}T$ and $^{13}K$ to the modified Korringa relation  
$1/^{13}T_{1}T=1/^{13}T_{1,orb}T+(^{13}K-^{13}K_{orb})^{2}/S\beta$ 
with $\beta=6.0$, the orbital contributions $1/^{13}T_{1,orb}T=0.005 
$ sec$^{-1}$K$^{-1}$, and $^{13}K_{orb}=0.014\%$.  When one is dealing with 
correlation effects in a three dimensional electron gas with a
spherical Fermi surface, $\beta=1$ for the uncorrelated case, and $\beta > 1$
for the ferromagnetically correlated case. On the other hand, the enhancement 
of low frequency spin 
fluctuations at non-zero wavevectors affects only 
$1/T_{1}T$, and hence tends to reduce $\beta$.  The 
successful fit of $1/^{13}T_{1}T$ and $^{13}K$ by a modified Korringa law 
with $\beta=6.0$ suggests that above 120K the d-p bands in MgCNi$_{3}$ 
form a Fermi-liquid state with relatively strong ferromagnetic correlation 
effects. This is consistent with the strong enhancement of the 
uniform spin susceptibility with decreasing temperature. 
We must caution, however, that $\beta=6.0$
may be somewhat too large due to the fact that we have ignored the 
presence of two bands that would tend to overestimate $\beta$ by a 
factor $\sim 2$ \cite{Yafet}.
Furthermore, as noted by Moriya\cite{Moriya}, the precise magnitude of $\beta$ 
 is sensitive to the deviation of the Fermi surface from spherical 
 symmetry. We call for more sophisticated band 
theoretical analysis of our data to clarify the nature of the 
correlation effects at the quantitative level.

Even though the modified Korringa law describes our data reasonably 
well above 120K, it is important to 
notice that a single modified Korringa relation cannot account for the 
continuous increase of $1/^{13}T_{1}T$ below 120K through $T^{*}\sim 
50$K down to $\sim20$K.  In this regime, $^{13}K$ shows a crossover to 
a low temperature constant regime.  The fact that the overall spin fluctuations 
reflected in the wave vector 
integral of $\chi''({\bf q},\omega_{n})$ increase while the ${\bf q}={\bf 0}$
 component of the total static spin suceptibility $\sum_{j}\chi_{j}'(q=0)$ maintains a 
constant magnitude strongly suggests that spin fluctuations with 
finite wave vectors away from ${\bf q}={\bf 0}$ continue to grow significantly 
below $T^{*}\sim 50$K.  In this context, it is important to realize 
that band calculations suggest the presence of strong nesting effects between some segments of 
the Fermi surface\cite{Band2,Band3}.  If the nesting effects are indeed substantial, spin 
fluctuations corresponding to the nesting wavevectors could indeed continue to grow 
below $T^{*}\sim 50$K.  

Both $^{13}K$ and $1/^{13}T_{1}T$ 
are saturated below 20K. Using the same orbital contributions as 
before, the modified Korringa relation for the correlated Fermi-liquid gives $\beta=4.7$ below 20K.  
This value is smaller than $\beta=6.0$ 
observed above 120K by 20\%, again signaling the importance 
below 20K of correlation effects with finite wave vectors.  
Putting all the pieces together, 
we obtain the following physical picture for the normal state of 
MgCNi$_{3}$: the electrons in the Ni-C d-p bands are modestly 
correlated; the primary 
channel of the correlation effects appear to be centered near ${\bf 
q}={\bf 0}$ above 50K but spin fluctuations with finite wave vectors ${\bf 
q}\neq{\bf 0}$
show continuous growth down to 20K below which electron correlation effects 
are saturated with R$_{W}$=1.15. Extensive efforts are underway to  
understand the electronic properties of MgCNi$_{3}$ based on band 
calculations \cite{Band1,Band2,Band3,Band4}
, and our experimental data provides a good testing ground for 
those theories. It is worthwhile recalling that a similar situation is also encountered 
in Sr$_{2}$RuO$_{4}$. The $^{17}$O NMR Knight shifts in Sr$_{2}$RuO$_{4}$ increase 
with decreasing temperature and even begin to decrease below $T^{*}\sim 
50$K. On the other hand, $1/^{17}T_{1}T$ continues to grow through 50K\cite{Imai}.
Subsequent inelastic neutron scattering measurements 
revealed that an anomalous enhancement of spin fluctuations corresponding to the nesting vectors 
of quasi-one dimensional 4d$_{yz,zx}$ bands is responsible for 
the continuous increase of $1/^{17}T_{1}T$ towards $T_{c}$\cite{Sidis}. 

Next, we turn our attention to the behaviour of $1/^{13}T_{1}$ in the superconducting 
state. As shown in Fig. 3a, a magnetic field of 9 Tesla destroys 
superconductivity down to $\sim$2.5K above which
$1/^{13}T_{1}$ maintains a Korringa behaviour $1/^{13}T_{1}T = 
0.072 $ sec$^{-1}$K$^{-1}$.
However, at lower magnetic fields, $1/^{13}T_{1}$ is enhanced just 
below $T_{c}$, peaks at $\sim 0.9 T_{c}(H)$, followed by an 
exponential decrease at lower temperatures. Moreover, the peak value of 
$1/^{13}T_{1}$ just below $T_{c}$ grows with decreasing magnetic field 
from 1 Telsa to 0.45 Tesla. For the lowest field at $H$ = 0.45 Telsa, 
$1/^{13}T_{1}$ is enhanced by a factor $\sim 1.4$ just below 
$T_{c}(H=0.45T) = 6.7$K.
A similar robust enhancement is usually observed in conventional BCS
superconductors due to the pile up of the density of quasi-particle 
excitations, and is known as the Hebel-Slichter coherence peak 
\cite{Hebel,Review}. The temperature dependence of the coherence peak 
can be fit by incorporating a minor {\bf k}-space anisotropy in the conventional 
s-wave gap \cite{Review}
\begin{equation}
\Delta (H,{\bf \Omega})  = <\Delta(H)> (1+a({\bf \Omega}))
\label{dell}
\end{equation}
where ${\bf \Omega}$ is the solid angle in {\bf k}-space, $<\Delta(H)>$ is the mean 
gap value over all orientations in {\bf k}-space and $a({\bf \Omega})$ 
is the small anisotropy function with the condition $<a({\bf 
\Omega})>=0$. The best fit results, shown in Fig. 
3, yield a mean gap magnitude of $<\Delta(H=0.45T)>/k_{B} \sim 10.5$K 
and mean square anisotropy $<a^{2}({\bf \Omega })
> \sim$ 0.047 in a magnetic field of 0.45 
Tesla, or equivalently $\frac {<\Delta (H)>}{k_{B} T_{c}(H)} 
\sim $ 1.6 which is consistent with the standard BCS value of $\frac 
{<\Delta(H=0)>}{k_{B} T_{c}(H=0)}$=1.75 \cite{Shift}. $<\Delta(H)>$ may be somewhat underestimated by 
the influence of $H$, as $<\Delta(H=1T)>$ is smaller than $<\Delta(H=0.45T)>$ 
as shown in Fig. 3b.

We also noticed that the exponential decrease of 
$1/^{13}T_{1}$ saturates below $\sim 2.5$K and that the magnitude of 
$1/^{13}T_{1}$ itself is distributed below this temperature. It is known \cite{Ong} that a 
large `flux avalanche' effect occurs below 3K in MgCNi$_{3}$, and the 
saturation of $1/^{13}T_{1}$ is most likely due to the fluxoid 
contribution \cite{Review,DeSoto,Stenger}. Unfortunately, the fluxoid 
effects prevented us from confirming the exponential decrease in 
$1/^{13}T_{1}$ down to the lowest temperature. 
We should caution that 
theoretically, a small peak in $1/^{13}T_{1}$ is known to 
arise even in non-s orbital pairing symmetries as long as 
singularities exist in the quasi-particle excitation spectrum. 
However, we are not aware of any
materials with anisotropic pairing 
symmetry that shows such a robust coherence peak with the appropriate 
magnetic field dependence.

Given the qualitative similarities of the normal state NMR data 
between MgCNi$_{3}$ and the unconventional 
superconductor Sr$_{2}$RuO$_{4}$\cite{Ishida}, the possible
realization of conventional s-wave symmetry superconductivity in the former may be rather surprising.
However, we point out that correlation effects in the three dimensional 
material MgCNi$_{3}$ are expected to 
be much weaker than in the quasi-two dimensional material Sr$_{2}$RuO$_{4}$ \cite{Band2,Band3}.  
Furthermore, the presence of the
light C atom in the structure might create high frequency phonons and 
help open the s-wave channel at a higher temperature than
the p- or d-wave channels. In 
fact, $T_{c}=7$K of MgCNi$_{3}$ is several times higher than $T_{c}=1.4$K of Sr$_{2}$RuO$_{4}$.

To summarize, we have reported the first $^{13}$C NMR measurements in 
MgCNi$_{3}$.  The strong temperature dependence of $^{13}K$ and 
$1/^{13}T_{1}T$ and subsequent saturation below $\sim 50$K and 
$\sim$20K respectively are 
consistent with the presence of modest electronic 
correlation effects, possibly assisted by nesting effects.  
This suggests that the 
electronic state has reached a modestly mass-enhanced Fermi-liquid 
like state prior to the superconducting transition. The robust coherence 
peak and subsequent exponential decrease of $1/^{13}T_{1}$ is 
consistent with s-wave pairing.

The work at MIT was supported by NSF DMR 98-08941 and 99-71264, and at Princeton 
by NSF DMR 97-25979 and DOE DE-FG02-98-ER45706. 


%
%

\begin{figure}
\caption{(a) A Fourier transformed $^{13}$C NMR spectrum (arb.
units) in MgCNi$_{3}$ obtained at 4.6K, 7K, and 300K in 1 Tesla.
(b) A typical $^{13}$C NMR spin echo recovery at 4.6K and 1 Tesla.
A 180-degree saturation pulse was employed. The solid line is the
best fit to a single exponential recovery with $T_{1}=3.3 $ sec.}
\label{Data}
\end{figure}

\begin{figure}
\caption{(a) Temperature dependence of the $^{13}$C NMR Knight 
shift $^{13}K$.  Solid curve is a guide for the eyes and dashed line
is $^{13}K=0.018\%$ for graphite. Insert to (a): 
$\sqrt{1/^{13}T_{1}T- c}$ vs $^{13}K$. The dashed line above 120K is 
with c = 0.005 sec$^{-1}$K$^{-1}$ and $\beta=6.0$ (see main
text). (b) Temperature dependence of $1/^{13}T_{1}T$ measured in 
9 Tesla. Solid curve is a guide for the eyes. Insert to (b): $1/^{13}T_{1}T$ below 150K.}
\label{Kfig}
\end{figure}

\begin{figure}
\caption{(a) $^{13}1/T_{1}$ measured at 0.45 Tesla ($\bullet$),
1 Tesla ($\circ$) and 9 Telsa ($\times$). Dashed line represents 
$^{13}1/T_{1}T = 0.072$ sec$^{-1}$K$^{-1}$. Solid line through 0.45 
Telsa data is the standard BCS fit with mean gap magnitude 
$<\Delta(H=0.45T)>/k_{B} \sim $10.5K and mean 
square anisotropy $<a^{2}({\bf \Omega })> \sim$ 0.047 (see text).
(b) Same data and lines as (a) plotted as R$_{s}$ = $1/^{13}T_{1}T$ divided by R$_{n}$=0.072 
sec$^{-1}$K$^{-1}$ on a semi-logarithmic scale as a function of inverse 
temperature $1/T$. Dashed-dot line through 1 Tesla data is a fit in 
the low temperature region to $e^{-<\Delta(H=1T)>/k_{B}T}$ with 
$<\Delta(H=1T)>/k_{B}\sim $8K.}
\label{1/T} 
\end{figure}

%
%


\begin{references}
\bibitem{Shift}Another conventional NMR approach to clarify the 
pairing symmetry of the superconducting state is to measure the Knight 
shift below T$_{c}$. Unfortunately, we found that the resonance 
linewidth becomes even broader than the magnitude of the normal state 
Knight shift due to fluxoid effects (see Fig. 1a).
\end{references}
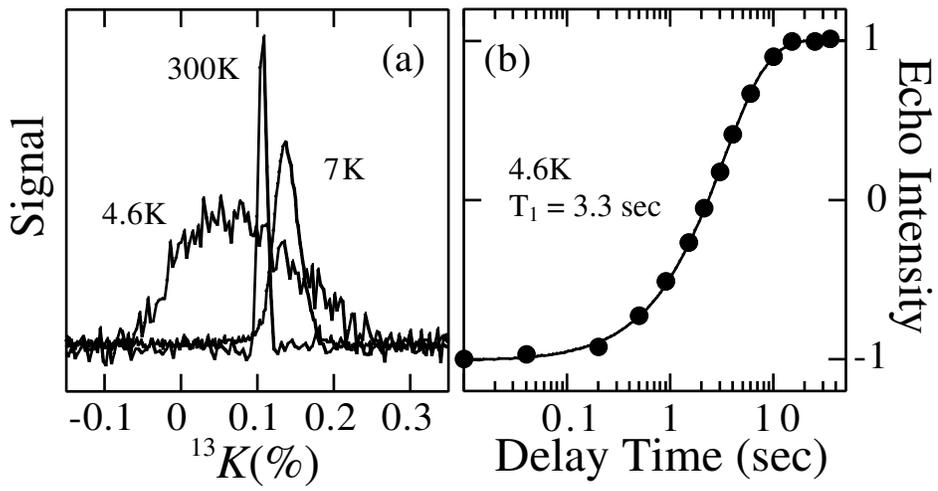
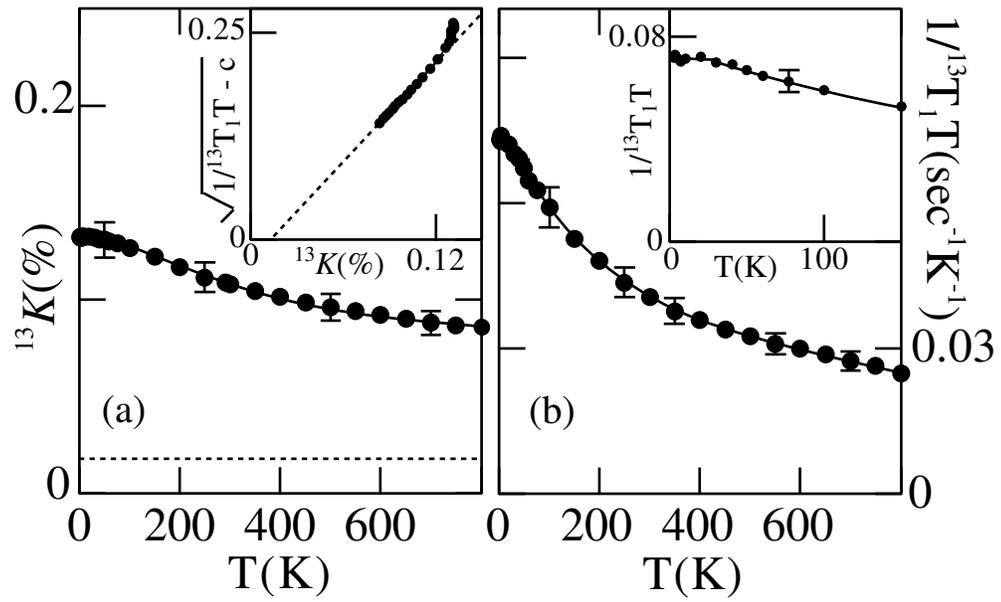
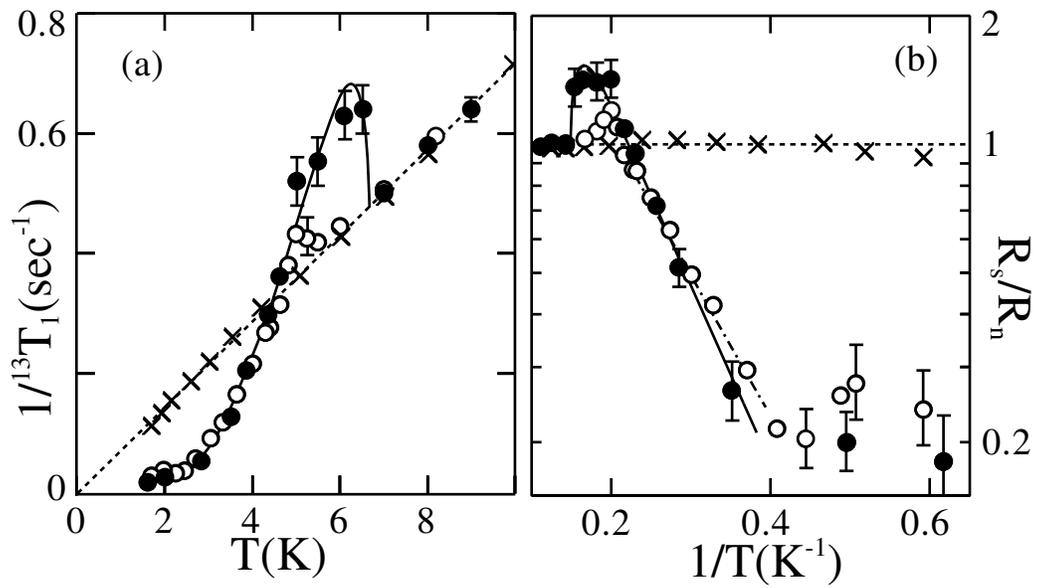
\end{document}